\documentclass[draft,preprint,superscriptaddress,showpacs,preprintnumbers,amsmath,amssymb,prb]{revtex4}

\usepackage{graphicx}
\usepackage{amsmath}
\usepackage{bbm}
\usepackage{verbatim}

\begin{document}

\preprint{}

\title{Tunnel spectroscopy in ac-driven quantum dot nanoresonators}

\author{J. Villavicencio}
\affiliation{Instituto de Ciencia de Materiales de Madrid (CSIC),
Cantoblanco, 28049 Madrid, Spain}
\affiliation{Facultad de Ciencias, Universidad Aut\'onoma de Baja
California, Ensenada, M\'exico}

\author{I. Maldonado}
\affiliation{Instituto de Ciencia de Materiales de Madrid (CSIC),
Cantoblanco, 28049 Madrid, Spain}
\affiliation{Centro de Investigaci\'on Cient\'ifica y de Educaci\'on
Superior de Ensenada, M\'exico}
\affiliation{Centro de Nanociencias y Nanotecnolog\'ia, Universidad
Nacional Aut\'onoma de M\'exico, Ensenada, M\'exico}

\author{R. S\'anchez}
\affiliation{Instituto de Ciencia de Materiales de Madrid (CSIC),
Cantoblanco, 28049 Madrid, Spain}

\author{E. Cota }
\affiliation{Centro de Nanociencias y Nanotecnolog\'ia, Universidad
Nacional Aut\'onoma de M\'exico, Ensenada, M\'exico}

\author{G. Platero}
\affiliation{Instituto de Ciencia de Materiales de Madrid (CSIC),
Cantoblanco, 28049 Madrid, Spain}

\date{published online 12 May 2008}

\begin{abstract}
Electronic transport in a triple quantum dot shuttle device in the
presence of an ac field is analyzed within a fully quantum
mechanical framework. A generalized density matrix formalism is used
to describe the time evolution for electronic state occupations in a
dissipative phonon bath. In the presence of an ac gate voltage, the
electronic states are dressed by photons and the interplay between
photon and vibrational sidebands produces current characteristics
that obey selection rules. Varying the ac parameters allows to tune
the tunneling current features. In particular, we show that coherent
destruction of tunneling can be achieved in our device.
\end{abstract}
%
%
%
\maketitle

The study of nanoelectromechanical systems (NEMS) has drawn a great
deal of interest, both at an applied and at a fundamental level, due
to novel transport regimes arising from the strong interplay between
electrical and mechanical degrees of freedom \cite{nems1}.
This is of particular interest in the field of molecular electronics
where, for example, the vibrational modes of a $C_{60}$ molecule are
coupled to its charge state giving rise to new features in the
current through the molecule \cite{Park}.

Shuttling is an example of nonlinear transport regime in NEMS, where
an oscillating nano-grain acts as a shuttle carrying one electron
from the left to the right contact.
A paradigmatic proposal of such a device was made in the seminal
work by Gorelik et al. \cite{gor}, and, since then, there have been
different models that describe the mechanical degree of freedom
using semiclassical \cite{pis01} or quantum mechanical
\cite{armour,cftnapjprb04} approaches.
In the presence of an ac electric field, it is known that the
interplay between  the ac and the natural frequency of the device
gives rise to rich dynamics with strong nonlinear characteristics.
These can be modified by tuning the ratio of the two frequencies or
the intensity of the applied field.
Up to now, this problem has been addressed for the case of an
oscillating quantum dot (QD) using a semiclassical approach to
describe the oscillations \cite{pis01}.

In this letter we introduce a full quantum mechanical description to
study the dynamics of a triple dot quantum shuttle (TDQS) with an ac
gate  voltage applied between the different QDs and including
quantum optical damping as a dissipation bath \cite{armour}.
Using the master equation approach for the reduced density matrix
(RDM), we show how tunneling current resonances occur through
different combinations of photonic and phononic sidebands and derive
sum rules to explain these resonances.
These features strongly depend on the ratio between the frequency
and intensity of the ac field.
Thus, by tuning the ac frequency we can obtain information on the
vibrational frequency of the resonator.
Finally, we find that the ac parameters can be manipulated so that
coherent destruction of tunneling \cite{gpra04PR,hang} (CDT) can be
observed.

%
%
We consider a TDQS, which consists of an array of three QDs where
the central movable dot is flanked by two static dots at fixed
positions $\pm x_0$.
%
The oscillation of the central dot will affect the tunneling rates,
which are now position dependent.
%
The charge transport is within the Coulomb blockade regime, so the
charging energy is assumed to be sufficiently large that only one
transport electron can occupy the chain of three dots at any given
time.
Our model introduces a time-dependent ac potential
$V(t)=V_{ac}\,\cos(\omega_{ac}t)$, applied between the left and
right QDs, where $V_{ac}$ and $\omega_{ac}$ are the amplitude and
frequency of the applied field, respectively (see Fig. \ref{driven
TDQS}).
The system is modeled by means of a Hubbard like Hamiltonian,
%
$\hat{H}=\hat{H}_{TQS}+\hat{H}_{leads}$,
where the unperturbed hamiltonian is
%
$\hat{H}_{TQS}=\hat{H}_0+\hat{H}_{osc}+\hat{H}_{tun}+\hat{H}_{ac}$,
where
$\hat{H}_0=\sum_{s} \varepsilon_{s}\,c_{s}^{\dagger }c_{s}$
represents the Hamiltonian of the left-hand ($l$), central ($c$),
and right-hand ($r$) dots, for spinless electrons with respective
localized states $\varepsilon_s$ ($s=l,c,r$).
The energy of the central QD is
$\varepsilon_c(\hat{x})=[\varepsilon_l-(\varepsilon_b/2x_0)(
\hat{x}+x_0)]$,
where it is assumed that it undergoes a Stark shift proportional to
its position due to the voltage bias across the device
$\varepsilon_b=(\varepsilon_l-\varepsilon_r)$;
$\hat{H}_{osc}=\hbar \omega d^{\dagger }d$ is the energy associated
to the oscillation of the central QD, with frequency $\omega $.
$\hat{H}_{tun}=\sum_{s'=l,r}T_{s'}(\hat{x})\,[c_c^{\dagger
}c_{s'}+h.c.]$, represents the tunneling between all three dots,
with position dependent tunneling rates
%
%
$T_l=-V \exp^{-\alpha(x_0+\hat{x})}$, and $T_r=-V
\exp^{-\alpha(x_0-\hat{x})}$,
where $V$ is the tunneling amplitude and $\alpha$ is the inverse of
the tunneling length.
The position operator $\hat{x}$ measures the displacement of the
vibrational mode and is given by
$\hat{x}=\Delta x_{zp}\,(\hat{d}^{\dag}+\hat{d})$,
where $\Delta x_{zp}=(\hbar/2m\omega)^{1/2}$ is the zero-point
uncertainty position of the oscillator.
$\hat{H}_{ac}=(V_{ac}/2) \cos ( \omega_{ac} t)\,( c_{l}^{\dagger
}c_{l}-c_{r}^{\dagger }c_{r})$, provides the effect of the ac
potential, which is introduced as an oscillation of opposite phase
on the energy levels of the right and left dots flanking the central
dot
and
$\hat{H}_{leads}=\sum_{\alpha}\sum_{k_{\alpha},\sigma}\epsilon_{k_{\alpha}}\,c_{k_{\alpha}
\sigma}^{\dagger }c_{k_{\alpha} \sigma}$.
The equation of motion of the system using the standard RDM approach
\cite{Blum}, generalized to include the environment of the
oscillator, is given by:
$\dot{\rho}_{qs}^{ij}=(\dot{\rho}_{qs}^{ij})_{RDM}
+(\dot{\rho}_{qs}^{ij})_{diss}$,
where $\rho_{qs}^{ij}$ represents the matrix element with the dot
states $q$ and $s$ ($q,s=0,l,c,r$), with the oscillator states $i$,
and $j$ ($i,j=0,1,2,...$). The term $(\dot{\rho}_{qs}^{ij})_{RDM}$
incorporates transitions between the leads and the outer dots:
\begin{eqnarray}
&&(\dot{\rho}_{qs}^{ij})_{RDM} =-\frac{i}{\hbar }[ \hat{H},\,\rho ]_{qs}^{ij} \nonumber \\
&&+\left\{
\begin{array}{cc}
\sum_{d\neq q}\left( \Gamma _{d\rightarrow q}\,\rho
_{dd}^{ij}-\Gamma
_{q\rightarrow d}\,\rho_{qq}^{ij}\right) ; \, q=s \\
-\frac{1}{2}\left( \sum_{d\neq q}\Gamma _{q\rightarrow
d}+\sum_{d\neq s}\Gamma _{s\rightarrow d}\right) \,\rho_{qs}^{ij};
\, q\neq s.
\end{array}
\right. \label{masterequation2}
\end{eqnarray}
Here we have also considered the transitions $\Gamma_{r\rightarrow
0}=\Gamma_{0\rightarrow l}=\Gamma$.
The term $[\dot{\rho}_{qs}^{ij}]_{diss}$
accounts for the dissipative effects of the oscillator's
environment. We assume a damping model composed of a bath of
oscillators at a fixed temperature $T$, to which the vibrational
mode is weakly coupled \cite{Scully}.
In the limit of small values of $T$, the matrix elements are given
by
$[\dot{\rho}_{qs}^{ij}]_{diss} =-\gamma
[(i+j)/2)]\rho_{qs}^{ij}+\gamma
[(i+1)(j+1)]^{1/2}\rho_{qs}^{i+1,j+1}$, where $\gamma $ is the
classical damping rate of the oscillator.
Finally, the matrix elements of the tunneling amplitudes
$T_{l,r}^{ij}$ in the oscillator basis can be computed.

%
%
In what follows, we shall explore the main features of the current
$I_{ac}$ for a TDQS driven by an ac gate voltage. We have considered
weak coupling between the three QDs, a regime characterized by large
values of $\alpha$.
The calculations are performed by numerical integration of the RDM
and averaging over the resulting electronic current measured at the
right QD, i.e., $I_{ac}=e \,\Gamma \,\sum_i
\,[\rho_{rr}^{ii}]_{av}$.
We begin our analysis by exploring $I_{ac}$ as a function of the
bias $\varepsilon_b$ for different values of the ac parameters.
All of the calculations are worked out in units where
$\hbar=2m=\omega=e=1$ and $N=6$ oscillator states were included.
%
For the particular values of the voltage bias across the device
$\varepsilon_b$ considered in our calculations, we have numerically
verified that including more oscillator states does not
significantly change the current characteristics for typical
experimental parameters.
%
%
%
In Fig. \ref{xigt1a}(a) we plot the time-averaged current $I_{ac}$
(solid line) for a particular value of the ratio
$\xi=(V_{ac}/2\omega_{ac})=3.0$, where $\omega=p\,\omega_{ac}$ ($p$
integer).
We can appreciate a much richer structure for $I_{ac}$ than the
undriven case (dashed line).
We find a series of peaks or maxima appearing at specific values of
$\varepsilon_b$, which are integer multiples of $\omega_{ac}$, and
obeying the following sum rules:
\begin{equation}
\varepsilon_b+\nu \hbar \omega_{ac}=n \hbar \omega, \label{sumrule1}
\end{equation}
\begin{equation}
(\varepsilon_b/2)+\nu \hbar \omega_{ac}=n \hbar \omega.
\label{sumrule2}
\end{equation}
Here, $n$ corresponds to the vibrational level at which the
transmitted electron ends up in the right dot, and $\nu$ to the
number of emitted or absorbed photons needed to reach that
vibrational level.
The peaks that satisfy the sum rule of Eq. (\ref{sumrule1}) are
positioned at a value of $\varepsilon_b$ equal to an \emph{odd}
multiple of $\omega_{ac}$, and are identified by $[ \nu, n ]$.
These correspond to an alignment between photon-assisted tunneling
(PAT) sidebands of the vibrational resonator states of the right and
left dots. As in the undriven case, these peaks are suppressed in
the very weak coupling regime. The peaks satisfying Eq.
(\ref{sumrule2}), positioned at an \emph{even} multiple of
$\omega_{ac}$, are identified with $(\nu,n)$, and correspond to
alignment of sidebands of all three dots.
In order to further corroborate the label assignments, we calculate
the occupation probabilities for the right QD $\rho_{rr}^{ii}$,
which determine $I_{ac}$. In Fig. \ref{xigt1a}(b), we present the
results corresponding to the peak at $\varepsilon_b\simeq 0.4$ [with
labels $(-1,0), (+4,1)$], where we can  clearly see that the most
important contribution arises from $\rho_{rr}^{00}$.
Similarly, for all the resonances we have verified that the most
important contribution comes from the satellite with the least
number of photons, which is consistent with a low field intensity.

In what follows we shall introduce a useful transformation in order
to demonstrate that the different peaks associated to the PAT
current are in fact associated with the emission or absorption of $
\nu$ photons.
%
%
Performing a unitary transformation \cite{rafa} $\hat{U}=\exp \left[
-i(V_{ac}/2\omega_{ac}) \sin ( \omega_{ac}t) (c_{l}^{\dag
}c_{l}-c_{r}^{\dag}c_{r})\right ]$  on the Hamiltonian $\hat{H}$,
leads to a transformed tunneling Hamiltonian
$\hat{H}^{\prime}_{tun}$ with time dependent tunneling amplitudes
$\tilde{T}_l=T_l\,\sum_{\nu =-\infty }^{\nu =+\infty }\,(-1)^{\nu
}\,J_{\nu }(\xi )\,e^{i\omega _{ac}\nu t}$, and $ \tilde{T}_r=
T_r\,\sum_{\nu =-\infty }^{\nu =+\infty }\,(-1)^{\nu }\,J_{\nu }(\xi
)\,e^{-i\omega _{ac}\nu t}$.
Thus, the tunneling amplitudes  become renormalized by a sum of
Bessel functions of order $\nu$, where $\nu$ is the number of
absorbed or emitted photons.
By computing the current using this approach we obtain an excellent
agreement with our previous calculation.
The peaks in $I_{ac}$ can be explained by an interplay of the
vibrational and photoassisted sidebands as follows:
the TDQS can be seen as a system of three fixed QDs, whose energy
levels are separated by ($\varepsilon_b/ 2$),
and the effect of the oscillator is to split these energies into
vibrational satellite sub-bands separated by the vibrational energy
$\hbar \omega$.
When the ac voltage is applied across the device, an additional
level splitting of their energies occurs by an amount of $\hbar
\omega_{ac}$, giving rise to new conduction channels (resonances)
which are accurately described by the sum rules.


We can now exploit the analytical properties of the Bessel functions
in order to control the current characteristics by an adequate
manipulation of the ac parameters.
In Fig. \ref{J_0} we use  $\xi=2.405$, for $\omega_{ac}=0.2$ and a
lower $V_{ac}=0.96192$, corresponding to the first zero of
$J_0(\xi)$.
In this case, direct tunneling processes involving $\nu=0$ at
$\varepsilon_b\simeq 0.0, 2.0$, are suppressed due to CDT
\cite{gpra04PR,hang}. Here, the energy levels of the vibrational
subbands of the three dots are aligned, but $\tilde{T}_l\approx0$
and $\tilde{T}_r\approx0$.
Note that these values of $\varepsilon_b$  also correspond to
resonances in the undriven case. Interestingly, the current peaks at
$\varepsilon_b\simeq 0.6, 1.0$, and $1.4$ [with labels $ [\nu, n]$],
observed in Fig. \ref{xigt1a}, are also significantly diminished.
This decrease in the current is due to two factors: the small
overlap of the wave functions in both left and right QDs and the
decrease in the intensity of the ac field, which yield a small
probability for processes involving several photons.
We see that the main peaks are governed by the emission and
absorption of several photons ($\nu=\pm4,..., \pm1$) and correspond
to alignment of the levels of all three dots.
%


In conclusion, we have analyzed the characteristics of the
electronic current in an ac driven TDQS device using a full quantum
mechanical approach.
%
%
%
Our results show that by tuning the ac field parameters we can
control the current contribution from differently vibrational modes.
This constitutes a new mechanism to obtain information on
vibrational modes through current measurements as for example in the
context of molecular shuttle devices.
%
%

We acknowledge A. Donarini for helpful discussions. This work has
been supported by the MCYT (Spain) under grant MAT2005-06444, and by
the EU Programme HPRN-CT-2000-00144. J.V. acknowledges support from
MEC-Spain: SB2005-0047. I.M. and E.C. acknowledge support from
CONACyT, M\'{e}xico, 43673-F.

\newpage

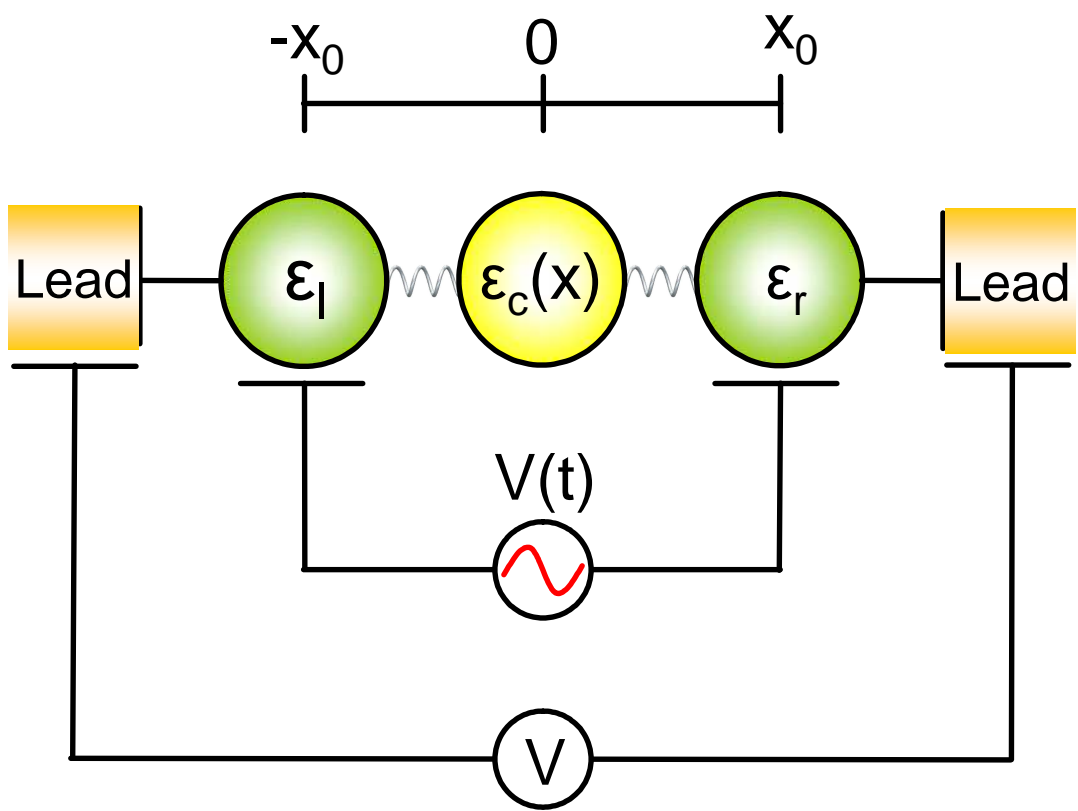
\begin{figure}[!h]
\caption{ Driven TDQS where oscillating tunneling barriers between
the central quantum dot and its neighbors are considered. The system
is driven by an ac potential $V(t)=V_{ac}\, \cos(\omega_{ac} t)$ and
the left and right tunneling barriers are rigidly coupled to the
leads.} \label{driven TDQS}
\end{figure}

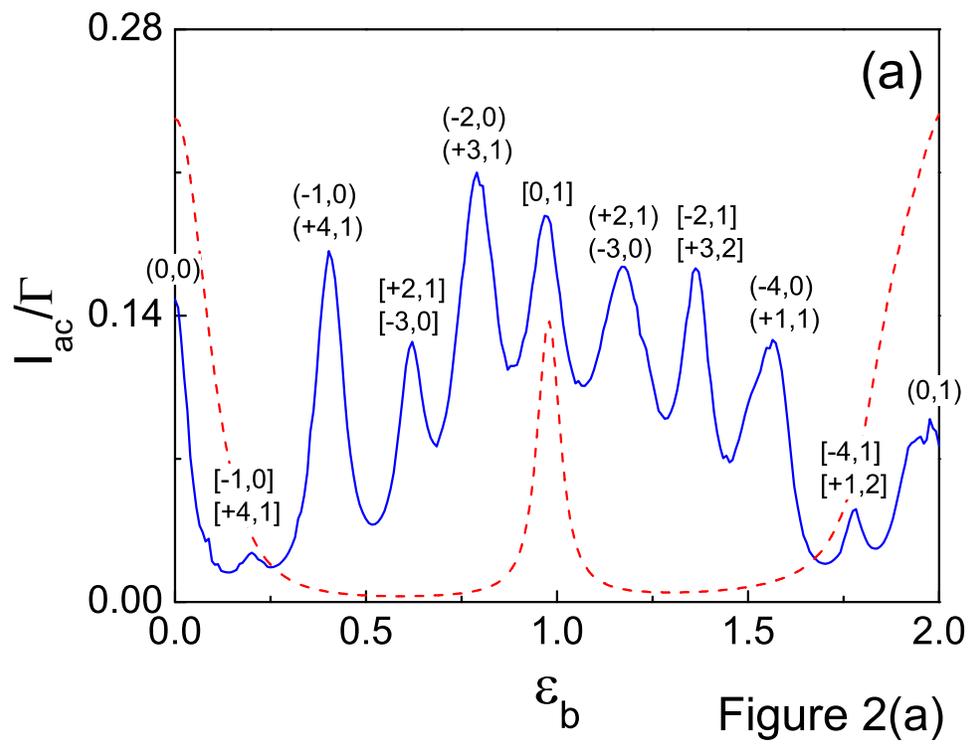
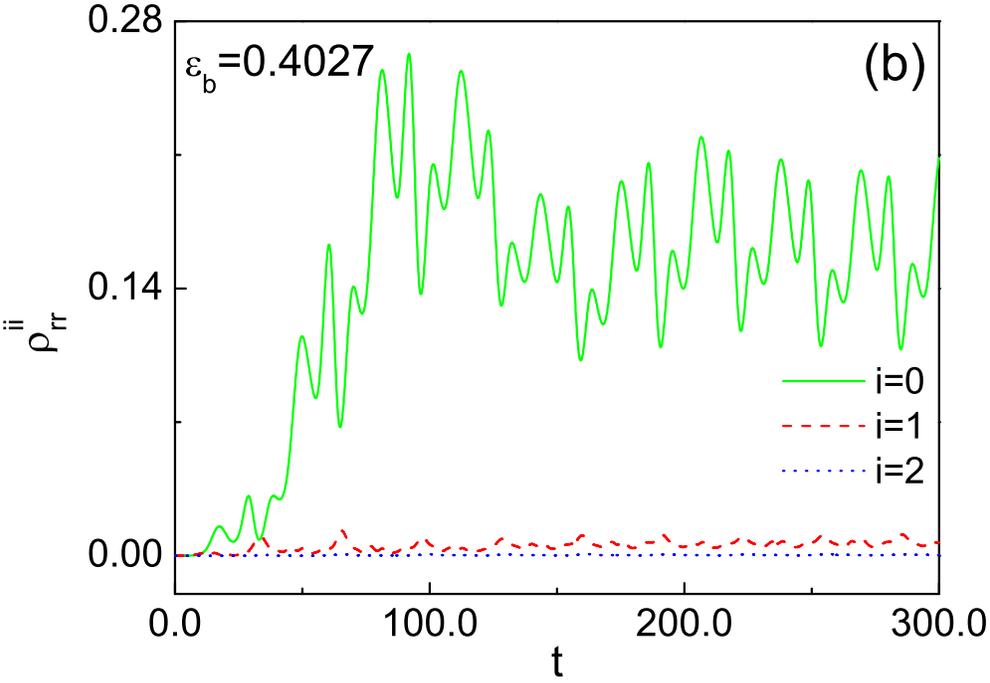
\begin{figure}[!h]
\caption{ (Color online) (a) $I_{ac}$ (in units of $\Gamma$) vs
$\varepsilon_b$ (blue solid line) for the ac parameters
$\omega_{ac}=0.2$, and $V_{ac}=1.2$, and typical values of the
parameters \{$V$, $\gamma$, $x_0$, $\Gamma$, $\alpha$\}=\{$0.5$,
$0.01$, $5.0$, $0.05$, $0.4$\}. The red dashed line corresponds to
$I_{ac}$ in the undriven case. The ac current peaks are labeled with
$[\nu,n]$, and $(\nu,n)$, associated to the sum rules given by Eqs.
(\ref{sumrule1}), and (\ref{sumrule2}), respectively. In (b) we show
the occupation probabilities of the right dot $\rho_{rr}^{ii}$, for
$i=0$ (green line), $i=1$ (red dashed line), and $i=2$ (blue dotted
line), as a function of time, corresponding to the current peak at
$\varepsilon_b\simeq 0.4$.} \label{xigt1a}
\end{figure}
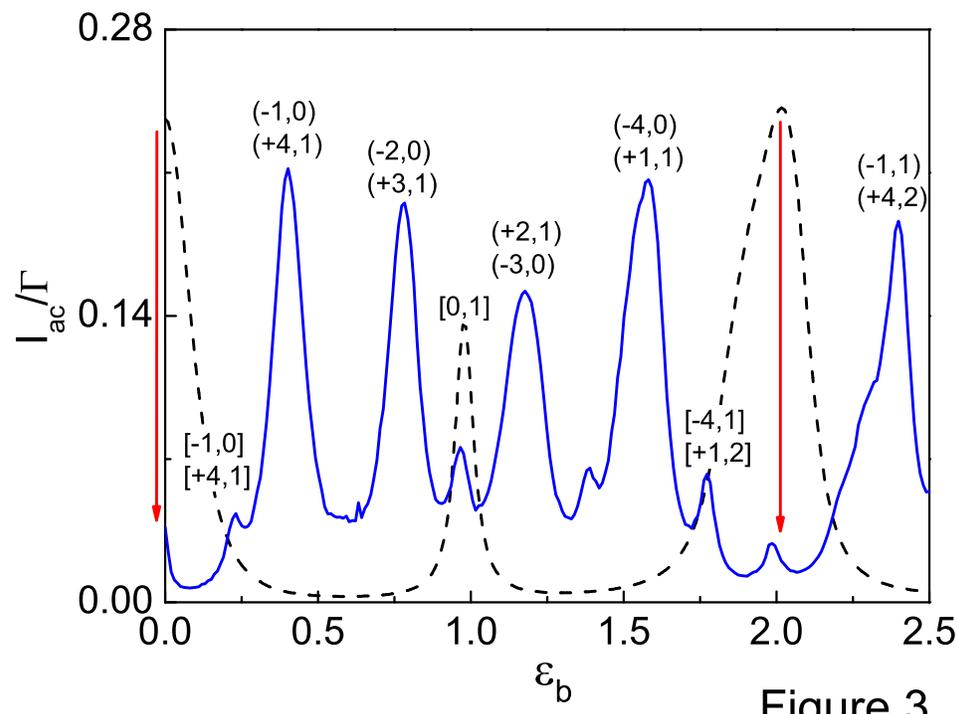
\begin{figure}[!h]
\caption{ (Color online) $I_{ac}$ (in units of $\Gamma$) as a
function of $\varepsilon_b$ (blue solid line), for
$\omega_{ac}=0.2$, $V_{ac}=0.96192$ ($\xi=2.405$), corresponding to
the first zero of $J_0(\xi)$, showing CDT. Parameters for the
undriven case (dotted line) are the same as in the previous figure.
} \label{J_0}
\end{figure}
\newpage

\end{document}